\documentclass[aps,prl,twocolumn,groupedaddress,showpacs]{revtex4}

\usepackage{graphicx}
\usepackage{bm}

\begin{document}

\title{Charge transfer, Symmetry and Dissipation}

\author{Maria R. D'Orsogna}
\author{Robijn Bruinsma}
\affiliation{Physics Department, University of California, Los Angeles, CA
90095-1547} 

\date{\today}

\begin{abstract}
We study charge transfer 
between donor-acceptor molecules subject to 
a mirror symmetry constraint in the presence of a dissipative environment.
The symmetry requirement 
leads to the breakdown of the standard single 
reaction coordinate description,
and to a new charge transfer theory, in the limit of low temperature,
based on two independent reaction coordinates of equal relevance.
We discuss implications of these results 
to charge transfer in DNA.
\end{abstract}

\pacs{82.20.-w, 82.20.Kh, 82.30.Fi, 82.39.Jn}

\maketitle

Charge trans\-fer be\-tween large or\-ganic molecules
in aqueous solvent plays a key role in biochemical
reactions, particularly those involving
animal and plant metabolism \cite{muller, chance}.
Compared to electronic transport in solid state materials,
electron transfer in proteins and DNA
is characterized by low values
of the tunneling matrix elements and strong coupling
between the electronic and nuclear degrees of freedom \cite{jortner}.
Charge transfer mainly occurs when the nuclear coordinates
happen to adopt a value for which the donor (D)
and the acceptor (A) state energies are
nearly degenerate.

In the conventional Born-Oppenheimer, or adiabatic, description,
such a degeneracy region transforms to a saddle-point
in the energy landscape separating the
D and A states, and the transfer rate 
reduces to the classical Kramers theory for activated chemical reactions
in a dissipative medium \cite{kramers, chandrasekhar, hanggi}.
However, because of the low value of the tunnel frequency 
$\Delta_0$ for biomolecular charge transfer, 
the adiabatic assumption is often not valid
and, as a result, a quantum-mechanical description 
is required to
determine the charge transfer rate.
The interaction of the nuclear degrees of freedom 
with the solvent medium plays an important role in this respect 
because not only does coupling to a finite-temperature heat
bath allow access to the degeneracy points by thermal activation,
but friction between the nuclear
degrees of freedom and 
the dissipative solvent - unavoidable due to the
fluctuation-dissipation theorem - increases the time spent by the 
nuclear degrees of freedom in the degeneracy region 
\cite{garg, marcus-sutin}.

The analytical theory of DA charge transfer
between biomolecules is based
on the assumption that, among the many 
nuclear degrees of freedom involved, there is always a single
dominant one: the \textit{reaction coordinate}
$Y$ that leads across the lowest saddle-point
in the adiabatic energy landscape separating 
the D and A states.
Charge transfer takes place
when $Y$ is within the Landau-Zener distance 
$l_{LZ} \propto \hbar \Delta_0 / \Delta F $
\cite{landau, nikitin}
of the degeneracy point, with $ \Delta F$
the typical force level
on the nuclear degrees of freedom 
at the degeneracy point \cite{note to DF}.
Cross-over to the adiabatic regime
occurs  when the time spent in the Landau Zener
region $\tau_{LZ}$ is large compared to the
typical tunneling time $\Delta_0^{-1}$.

Single reaction DA theory is a basic tool
not only for describing donor-acceptor 
charge transfer processes but also for 
describing chemical reactions 
involving macromolecules in general.
This single reaction
coordinate theory is assumed to be
at least qualitatively correct, 
though relaxation of the remaining nuclear degrees of freedom
at the transition state may renormalize the
parameters of the theory. 
It is the aim of the present letter to 
demonstrate that the single coordinate description
fails in the presence of symmetry constraints.
To be specific, consider charge transfer between
two identical molecules. The electronic degree
of freedom will be represented by Pauli spin matrices
with $\sigma_z=1$ denoting the electron in the
D state and $\sigma_z=-1$ in the A state.
The electron is coupled to the same 
two nuclear degrees of freedom of the D and A 
molecule $Y_1$, respectively, $Y_2$.
The Hamiltonian is:

\begin{eqnarray}
\label{Hamiltonian}
H &=&\frac{P_{Y_1}^2}{2M} + \frac{P_{Y_2}^2}{2M} + V(Y_1, Y_2, \sigma_z)
+\frac{\hbar \Delta_0}{2} \sigma_x +\\
&&
\nonumber
\sum_{\alpha, ~ i=1,2} \left[ \frac{p_{\alpha,i}^2}{2m_\alpha}
+\frac 1 2 m_\alpha \omega_{\alpha}^2
\left( x_{\alpha,i}+ c_\alpha \frac{Y_i}{m_\alpha \omega_{\alpha}^2} 
\right)^2
\right]
\end{eqnarray}

\noindent
Here, $M$ is the effective mass and $V$ the potential energy
of the two nuclear degrees of freedom.
Two collections $\{ x_{\alpha_i}\}$ of harmonic
oscillators 
represent the environmental degrees of freedom.
They are coupled separately to the nuclear degrees of freedom, 
and generate
a frictional drag on $Y_1$ and $Y_2$
with a friction constant
\cite{leggett}:

\begin{equation}
\label{dissipation}
\eta = \lim _{\omega \to 0} \frac \pi {2 \omega} \sum_{\alpha}
\frac{c_{\alpha}^2}{m_\alpha \omega_\alpha} 
\delta(\omega - \omega_{\alpha}).
\end{equation}

\noindent
In the absence of any symmetry constraints,
the lowest order coupling between the nuclear and
electronic degrees of freedom is of the form 
$(Y_1 - Y_2) \sigma_z$. 
Treating $(Y_1 - Y_2)$ as the reaction coordinate,
we can apply the standard single coordinate formalism.
However, if we impose a mirror symmetry
$Y \rightarrow -Y$, $\{x_\alpha \rightarrow -x_\alpha \}$
then this term is forbidden. 
Expanding the potential energy to the lowest order
in the lowest order in the nuclear coordinates
under the symmetry constraint
gives a Landau-Ginsburg type potential:

\begin{eqnarray}
\label{potential}
\nonumber
V(Y_1,Y_2,\sigma_z) &=& \frac 1 2 k (Y_1^2+Y_2^2)+\frac 1 4 
v (Y_1^4 + Y_2^4) - \\
&&
\frac{\lambda k}{4}
\left[ (1+\sigma_z) Y_1^2+ (1-\sigma_z) Y_2^2 \right].
\end{eqnarray}

\noindent
Here, $k$ is a spring constant, $v$
describes the stabilizing effect of the lowest order
anharmonic term, and $\lambda$
is the dimensionless coupling constant between
electronic and nuclear degrees of freedom.

We restrict ourselves to the case $1<\lambda<2$.
In this regime, the donor state energy surface 
$\left< \uparrow | V | \uparrow \right>$
has two minima, at
$Y_1=\pm \sqrt {k (\lambda -1) /v}, ~ Y_2=0$,
corresponding to a charge deformed
molecule in one of two alternative
mirror related structures.
We denote the `left' and `right' 
donor structures respectively
by $|$ L $ \left. \uparrow \right>$ 
and $|$  R $ \left. \uparrow \right>$.
The acceptor state energy surface 
$\left< \downarrow | V | \downarrow \right>$
has two corresponding minima but rotated 
over $\pi/2$ in the $(Y_1, Y_2)$
plane, as shown in Fig. 1.
To locate the degeneracy points,
we must solve for $\left< \uparrow | V | \uparrow \right> =
\left< \downarrow | V | \downarrow \right>$,
which yields two degeneracy lines
$Y_1= \pm Y_2$ that cross at the origin
of the ($Y_1$, $Y_2$) plane.
The lowest degeneracy point,
the putative transition state,
is the origin with an energy barrier
$\Delta E = \frac 1 4  k (\lambda -1)^2/v$.

\begin{figure}
\includegraphics[height = 2.5 in]{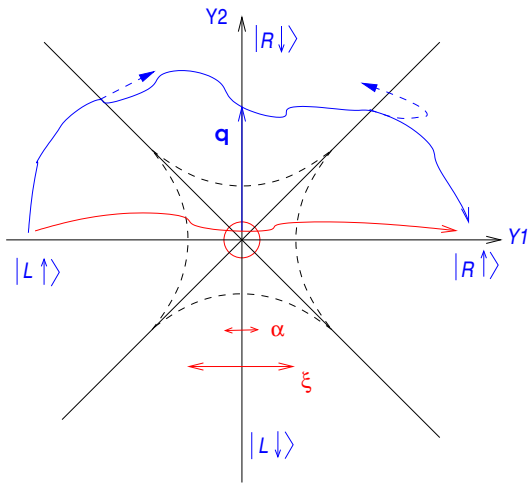}
\caption
{Figure illustrating two different reaction paths
in the $Y_1, Y_2$ plane of the nuclear coordinates. The locus of degeneracy
points between the donor and the acceptor state energies are indicated by the
bold lines $Y_1 = \pm Y_2$. The Landau-Zener region, within which 
charge-transfer reactions can take place, is shown as a dashed line. The 2D
Landau-Zener length $\xi$ defines a region
surrounding the origin where 
single-reaction coordinate theory fails. At the center of the 2D region one
encounters strong resonant tunneling within a distance $\alpha$ from the 
origin. The top trajectory, with a large 
impact parameter $q$, crosses the two degeneracy
lines separately. 
Charge-transfer at the crossing points allows transfer to the
acceptor state $\left. 
|R \downarrow \right>$ (dashed arrows). The bottom 
trajectory, with a low impact parameter, 
enters both the 2D region and the coherent tunneling region.}
\label{fig:lzener1}
\end{figure}

To see why a single reaction coordinate
description is not consistent in the presence of 
the symmetry constraint, we show in Fig. 1
two possible trajectories in the $(Y_1, Y_2)$
plane for a thermally activated 
hop in the donor state from the L to the R configuration.
The trajectories are classified by the impact parameter
$q$, the distance of closest approach to the origin.
A typical trajectory will cross each 
of the two degeneracy lines at least once.
At each crossing point a charge transfer
event can take place, with the system 
eventually ending in one of the
two acceptor minimum states.
Within a single reaction coordinate description 
for the charge transfer events, each degeneracy line is at the center of a 
Landau-Zener
region of width $l_{LZ}(q) \propto \hbar \Delta_0 / k \lambda q$.
For low impact parameters, the two 
Landau-Zener regions of the degeneracy lines
start to overlap when $l_{LZ}(q)$
drops below $q$, i.e. 
when $q$ is less than $\xi = \sqrt{\hbar \Delta_0/ k \lambda}$.
As a result, there is a region of size 
$\xi$ surrounding the origin 
where charge transfer is inherently 
\textit{two dimensional}.
If the thermal energy is low compared to the barrier height
$\Delta E$, then this 2D
region will dominate the charge transfer reaction and the single 
reaction coordinate
assumption is not valid.

In order to analyze a 2D charge transfer event
we generalize the Smoluchowski-Zusman (SZ)
method \cite{garg, zusman} to two dimensions.
In principle, 
this method is restricted
to the case of strong damping, when
the thermal energy $k_{B} T$
is large compared 
to $\hbar \omega_c$, with $\omega_c$ the
relaxation rate of the reaction coordinates. 
For the 1D case though, the SZ method 
reproduces the weak coupling non-adiabatic result 
in the limit of low tunneling rates.

In a 2D SZ description, the $2$ x $2$
density matrix $n_{i,j}$
of the ``spin'' degrees of freedom
($i,j$ now indicate spin up and down respectively) 
obeys the following transport equation:

\begin{eqnarray}
\label{SZ1}
\frac{\partial n_{i,j}}{\partial t} &=&
\frac 1 \eta \vec{\nabla} \left\{ 
\vec{F_{i,j}}+ k_B T ~ \vec{\nabla} n_{i,j}
\right\}
-\frac{i}{\hbar}
[H_\sigma, n_{i,j}],
\\
\label{SZ2}
H_{\sigma} &=& \frac{\hbar \Delta_0}{2} \sigma_x
-\frac{\lambda k}{4} [Y_1^2-Y_2^2] \sigma_z,
\end{eqnarray}

\noindent
with $\vec{F_{i,j}} = -\vec{\nabla}
\left< i | V | j \right>$
the $2$ x $2$ force matrix derived 
from Eq.(\ref{potential}).
The first term on the right hand side of 
Eq. (\ref{SZ1})
is the Smoluchowski operator for the
2D diffusive motion of a classical
particle in a force field.
The second term describes the precession of a spin 1/2
degree of freedom with the spin Hamiltonian $H_\sigma$
given by Eq. (\ref{SZ2}).
Note that the energy difference between spin up and 
spin down has a saddle point at the origin.
The characteristic length scale appearing in 
$H_\sigma$ is
the 2D Landau-Zener length
$\xi$ encountered above.
A comparison of $H_\sigma$
with the Smoluchowski
operator produces the second length
scale $\alpha_{2D} = (\hbar k_B T / \eta k) ^ {\frac1 4}$,
which is the characteristic variation length of
$n_{1,2}$. 
In the regime of strong damping, $\alpha_{2D}$ 
is small compared to $\xi$.

The equation for the off-diagonal part
of the density matrix can be
solved under the condition that $n_{1,2}$
varies over length-scales
that are short compared to those 
of the diagonal terms $n_{1,1}$ and $n_{2,2}$:

\begin{equation}
\label{imaginary}
Im ~ n_{1,2} (\vec{Y},t) \simeq 4 \sqrt 2 \pi \xi^2 ~ [n_{1,1}(0,t) - 
n_{2,2} (0,t)] ~ \delta(\vec{Y}).
\end{equation}

\noindent
The 2D delta function in Eq. (\ref{imaginary})
for the nuclear degrees of freedom is actually a 
gaussian with 
a width of order $\alpha_{2D}$. Inside this gaussian
region, the off-diagonal part of the density matrix is 
large: the spin degree of freedom 
is precessing coherently. Physically,
this means that resonant tunneling
is taking place between the D and A
states within a distance $\alpha_{2D}$ of the origin of
the ($Y_1$, $Y_2$) plane.
Substituting Eq.(\ref{imaginary})
in Eq.(\ref{SZ1}), we find that the diagonal terms of the 
density matrix obey a classical
Smoluchowski
equation with a ``sink'' at the origin:

\begin{eqnarray}
\frac{\partial n_{1,1}}{\partial t} &=& \frac 1 \eta
\vec{\nabla} \left\{\vec{F}_{11}+k_B T \vec{\nabla}
\right\} n_{1,1} - \\
\nonumber
&&
4\pi \sqrt 2 \Delta_0 \xi^2
(n_{1,1}-n_{2,2}) ~ \delta(\vec{Y}),
\label{density1} \\
\nonumber
\\
\frac{\partial n_{2,2}}{\partial t} &=& \frac 1 \eta
\vec{\nabla} \left\{\vec{F}_{22}+k_B T \vec{\nabla}
\right\} n_{2,2} + \\
\nonumber
&& 
4\pi \sqrt 2 \Delta_0 \xi^2
(n_{1,1}-n_{2,2}) ~ \delta(\vec{Y}).
\label{density2}
\end{eqnarray}

\noindent
The decay rate of the donor state can be computed from
Eq. (\ref{density1}) and Eq. (\ref{density2}) 
using standard methods:

\begin{equation}
\label{decay}
\frac{\Gamma}{\Gamma_0} = 4
\frac{\sqrt{2(\lambda-1)}}{\lambda} 
\displaystyle{\frac{\displaystyle{\frac{\hbar \Delta_0^2}{ \omega_c k_B T}}}
{1 + C ~ \displaystyle{
\frac{\sqrt{\lambda-1}}{\lambda}} ~
\frac {\hbar \Delta_0^2}{\omega_c k_B T}}}.
\end{equation}

\noindent
Here, $\omega_c^{-1} = \eta / k (\lambda-1)$
is the classical life-time of the transition state,
and $\Gamma_0$ is the classical Kramers
rate for an activated hop to the transition state.
The constant $C$ equals $4 \sqrt 2 \ln 2$.

The 2D decay rate of Eq.($\ref{decay}$) 
is in the same form of the $1D$ 
description for donor-acceptor 
charge transfer provided  
we interpret $g_{2D} = \hbar \Delta_0^2/ \omega_c k_B T $
as the new adiabaticity parameter.
For a 1D charge transfer event, the adiabaticity parameter
is of the form $g_{1D} \propto \hbar \Delta_0^2/ \omega_c \Delta E_b$
with $\Delta E_b$ the characteristic energy scale of the nuclear 
degrees of freedom, such as the activation barrier.
In the low temperature limit $k_B T <\!< \Delta E_b$,
the effective 2D adiabaticity parameter
diverges while $g_{1D}$ remains finite. 
Physically, this means that in the low temperature 
limit, the adiabatic description is \textit{always} valid
in the presence of the symmetry constraint.

We can justify treating $g_{2D}$ as an adiabaticity parameter
by estimating the time 
$\tau_{2D}$ spent 
in the 2D Landau-Zener region during a hop event.
By Einstein's relation,
the classical diffusion constant $D$
of the nuclear degrees of freedom is $k_B T / \eta$,
so that $\tau_{2D} \sim \xi^2 / D$
is of order $ \xi^2 \eta / k_B T$. 
Using $\xi = \sqrt{\hbar \Delta_0/ k \lambda }$
and identifying $\Delta_0 \tau_{2D}$
as the adiabaticity parameter
we recover the above expression for $g_{2D}$.
Interestingly, even though the quasi-classical 
method should become generally valid in the low temperature 
limit due to the symmetry constraint, 
the latter enhances the resonant tunneling regime as well.
In the 1D description, resonant tunneling
takes place within a distance 
$\alpha_{1D} = (k_B T \hbar/\eta \Delta F)^{1/3} $
of the degeneracy point.
Compared to 
$\alpha_{2D} = (k_B T \hbar/\xi k)^{1/4} $
for the present case,
we see that for $T \to 0$ the regime of coherent tunneling
in $2D$ is always larger than in 1D.

We conclude by noting that, a DA charge transfer 
system can be viewed
as a two-state quantum system coupled
to quasi classical macroscopic degrees of
freedom (the nuclear coordinates).
Systems of this type form the basic elements
for quantum computation, the ``qubits''.
For our result to be of any interest in this context,
the adiabaticity parameter
$\Delta_0 \tau_{LZ}$ must be very large compared to one, of
the order of $10^4$ or larger \cite{qubit}.
Since the symmetry constraint leads to 
the divergence of the adiabaticity parameter
in the low temperature limit, DA systems
obeying such a constraint would be better suitable
for that purpose.
A possible realization of the symmetry constraint
is encountered for charge transfer between two 
DNA bases.
Molecular dynamics simulations
by Chen \textit{et al.} \cite{chen}
report that the 
in plane mirror symmetry of individual DNA
bases is broken when a hole occupies the
purine base. The mirror symmetry for
stacked bases along a B DNA chain is not exact
due to the propeller twist
but the two alternative L and R
structures of a charged base still may be sufficiently close
for the present model to be relevant.
It would be very interesting to test whether
charge transfer events
along DNA correlate with thermally activated hops 
between the L and R states.



\begin{references}

\bibitem{muller} \textit{Electron and Proton Transfer in Chemistry and 
Biology}, edited by A. Muller \textit{et al.}
(Elsevier, New York, Amsterdam, 1992).

\bibitem{chance} \textit{Tunneling in Biological Systems}, 
edited by B. Chance \textit{et al.}
(Academic, New York, 1980).

\bibitem{jortner} \textit{Electron Transfer from Isolated 
Molecules to Bio\-mole\-cules}, 
edited by J. Jortner and M. Bixon 
(Wiley, New York,  1999).

\bibitem{kramers} H. A. Kramers, Physica (Utrecht)
 \textbf {7}, 284 
(1940).

\bibitem{chandrasekhar} 
S. Chandrasekhar, Rev. Mod. Phys. \textbf{15}, 3 
(1943).

\bibitem{hanggi} 
P. Hanggi, P. Talkner and M. Borkovec,
Rev. Mod. Phys. \textbf{62}, 
2 (1990).

\bibitem{garg}
    A. Garg, J. Onuchic, and V. Ambegaokar, J. Chem.  Phys.  \textbf{ 83},
    4491 (1985).

\bibitem{marcus-sutin}
    R. A. Marcus and N. Sutin, Biochim.  Biophys.  Acta \textbf{ 811}, 25
    (1985).

\bibitem{landau}
    L. D. Landau and E. M. Lifshitz, \textit{Quantum Mechanics}, $3^{rd}$
    ed. (Pergamon, New York, 1977).

\bibitem{nikitin}
    \textit{Chemische Elementarprozesse}, edited by J. Heidberg 
    \textit{et al.} (Springer-Verlag, Berlin, New York, 1968).

\bibitem{note to DF} More precisely, $\Delta F$ is the difference between the
forces at the degeneracy point.



\bibitem{leggett}
    A. O. Caldeira and A. J. Leggett, Ann.  Phys.  (N.Y.) \textbf{ 149},
    374 (1983).

\bibitem{zusman}
   L. D. Zusman, Chem. Phys. \textbf{49}, 295 (1980).

\bibitem{chen}
   E. S. Chen and E. C. Chen, Biochem. Biophys. Res. Comm. 
	\textbf{289}, 421 (2001).

\bibitem{qubit}
   M. Nielsen and I. Chuang, \textit
	{Quantum Computation and 
   Quan\-tum In\-for\-ma\-tion}
   (Cambridge University Press, Cambridge, 2000).

\end{references}
\end{document}